\documentclass[12pt,english]{article}
\usepackage[T1]{fontenc}
\usepackage[latin1]{inputenc}
\usepackage{a4wide}
\usepackage{graphicx}
\usepackage{amssymb}

\makeatletter

\providecommand{\tabularnewline}{\\}

\usepackage{bbm}
\usepackage{amsmath}

\usepackage{babel}
\makeatother
\begin{document}
\setcounter{page}{0}

$\,\,$
\vspace{1cm}

\begin{center}\textbf{\LARGE $\mathbb{Z}_{2}$ monopoles in $SU(n)$
Yang-Mills-Higgs theories}\end{center}{\LARGE \par}
\vspace{1cm}

\begin{center}\textbf{Marco A.C. Kneipp}%
\footnote{kneipp@fsc.ufsc.br%
} \textbf{and Paulo J. Liebgott}%
\footnote{pj.liebgott@gmail.com%
}\end{center}

\begin{center}{\em Universidade Federal de Santa Catarina (UFSC),\\ 

Departamento de F\'\i sica, CFM,\\

Campus Universit\'ario, Trindade,\\

88040-900, Florian\'opols, Brazil.  \\ }

\end{center}

\vspace{0.3cm}

\begin{abstract}
$\mathbb{Z}_{n}$ monopoles are important for the understanding of
the Goddard-Nuyts-Olive duality when the scalar field is not in the
adjoint representation. We analyze the $\mathbb{Z}_{2}$ monopole
solutions in $SU(n)$ Yang-Mills-Higgs theories spontaneously broken
to $Spin(n)/\mathbb{Z}_{2}$ by a scalar in the $n\times n$ representation.
We construct explicitly a $\mathbb{Z}_{2}$ monopole asymptotic form for each of the weights of the defining representation of
the dual algebra $so(n)^{\vee}$.

\vfill 

\thispagestyle{empty}
\end{abstract}
\newpage

\section{Introduction}

Electromagnetic duality in Yang-Mills-Higgs theories was proposed
in the work of Goddard, Nuyts, and Olive (GNO) \cite{GNO}. In this
work they consider a theory with gauge group $G$ and a scalar field
$\phi$ in an arbitrary representation which spontaneously breaks
$G$ to $G_{0}$ in such a way that $\pi_{2}(G/G_{0})$ is nontrivial,
which is the necessary condition for the existence of topological
monopoles. Since then, the monopole's solutions\cite{GoddOliveReview,Bais1,Weinberg papers,Goddard and Olive papers}
and the electromagnetic duality conjecture\cite{Montonen-Olive Duality}
were studied intensely in the particular case where the scalar field
was in the adjoint representation. In this case, the unbroken gauge
group $G_{0}$ necessarily has a $U(1)$ factor which is generated
by the scalar field vacuum solution. This $U(1)$ factor guarantees
that $\pi_{2}(G/G_{0})=\mathbb{Z}$, in which case the theory can
have monopole solutions which are a generalization of 't Hooft-Polyakov
solution. We shall call them $\mathbb{Z}$ monopoles. The Bogomol'nyi-Prasad\_Sommerfield
(BPS) $\mathbb{Z}$ monopoles are conjectured to be dual to particles
in the adjoint representation in the dual theory for suitable supersymmetric
theories\cite{Montonen-Olive Duality}. On the other hand, much less
is known in the cases where $\phi$ is not in the adjoint representation
and $G_{0}$ is semisimple. In these cases, since $\pi_{2}(G/G_{0})$
is a cyclic group $\mathbb{Z}_{n}$ or a product of cyclic groups,
the monopoles are called $\mathbb{Z}_{n}$ monopoles. Some general
properties of these $\mathbb{Z}_{n}$ monopoles were analyzed in \cite{GNO,WeinbergZ_2mono,BaisZ_2mono},
and in \cite{StrasslerZ_2} it was proposed that the $\mathbb{Z}_{2}$
monopole in ${\cal N}=1$ $SU(n)$ super Yang-Mills should satisfy
a duality transformation which is alternative to the GNO conjecture. 

One of the main motivations for the study of monopoles and electromagnetic
dualities is their possible application to the problem of confinement
in QCD. Following the ideas of 't Hooft and Mandelstam, the confinement
of particles in QCD must be a phenomenon dual to confinement of monopoles
in a superconductor. In this model, the formation of chromoelectric
flux tubes in QCD must be due to a monopole condensate. However, it
is not clear yet if this condensate is made of $\mathbb{Z}$ monopoles,
$\mathbb{Z}_{n}$ monopoles, or Dirac monopoles. There are some lattice
results which indicate that confinement could be related to $\mathbb{Z}_{n}$
monopoles' condensates\cite{Halliday and Schwimmer}. Another more
recent application of GNO duality is in the geometric Langlands program
\cite{langland}.

In the last years, the ideas of  't Hooft and Mandelstam were applied
to supersymmetric non-Abelian theories which satisfy electromagnetic
duality with $\mathbb{Z}$ monopoles. In particular in \cite{Z(N) strings}
the confinement of $\mathbb{Z}$ monopoles by the formation of magnetic
flux tubes or $\mathbb{Z}_{n}$ strings in soft broken ${\cal N}=4$
super Yang-Mills theories with an arbitrary simple gauge group was
analyzed. It was shown that the tensions of these $\mathbb{Z}_{n}$
strings satisfy the Casimir scaling law in the BPS limit, which is
believed to be the behavior that the chromoelectric flux tubes in
QCD must satisfy. This result indicates that these $\mathbb{Z}_{n}$
strings can be dual to QCD chomoelectric strings.

In order to understand better the properties of the $\mathbb{Z}_{n}$
monopoles, in the present work we obtain explicitly the asymptotic
form of the $\mathbb{Z}_{2}$ monopoles in $SU(n)$ Yang-Mills-Higgs
theories with the gauge group broken to $Spin(n)/\mathbb{Z}_{2}$
by a scalar in the $n\times n$ representation of $SU(n)$ or its
symmetric part. In this case, one could in principle embed the theory
in a (deformed) ${\cal N}=2$ $SU(n)$ super Yang-Mills with a hypermultiplet
in the $n\times n$ representation which has vanishing $\beta$ function,
similarly to ${\cal N}=4$ super Yang-Mills theory. 

In the work of GNO, they obtained that the possible magnetic weights
of the $\mathbb{Z}_{n}$ monopoles must belong to the weight lattice
$\Lambda_{\omega}(\widetilde{G}_{0}^{\vee})$ of the dual unbroken
gauge group $\widetilde{G}_{0}^{\vee}$, where $\widetilde{G}_{0}$
means the covering group of $G_{0}$. The magnetic weights satisfy
a further constraint to belong to particular cosets in $\Lambda_{\omega}(\widetilde{G}_{0}^{\vee})$.
In order to obtain that constraint we used the fact that for a theory
with an unbroken gauge group $G_{0}=\widetilde{G}_{0}/K$, the $\mathbb{Z}_{n}$
monopole's topological charge sectors are associated to the elements
of the group $K$ which is a subgroup of $Z(\widetilde{G}_{0})$,
the center of the group $\widetilde{G}_{0}$. Then, we used the result
that the elements of $Z(\widetilde{G}_{0})$ are associated to cosets
which are related to nodes of the extended Dynkin diagram of $\widetilde{G}_{0}^{\vee}$
related to the node $0$ by a symmetry transformation. Therefore,
the $\mathbb{Z}_{n}$ monopoles must be associated to weights of a
subset of these cosets. This form of writing elements of the center
of a group was also used to obtain $\mathbb{Z}_{n}$ strings solutions\cite{Z(N) strings,casimir law,string integrability}.
Then, using the elegant general construction of Weinberg et al.\cite{WeinbergZ_2mono}
for the monopoles in theories where the gauge group $SU(n)$ is broken
to $Spin(n)/\mathbb{Z}_{2}$, we associated to each weight of the
defining representation of the dual algebra $so(n)^{\vee}$ a $su(2)$
subalgebra and constructed explicitly a $\mathbb{Z}_{2}$ monopole
asymptotic solution which we called fundamental, generalizing the
$\mathbb{Z}_{2}$ monopole solution for the $SU(3)$ gauge group \cite{WeinbergZ_2mono}.
This is consistent with the result in \cite{BaisZ_2mono} where the
authors concluded that the $\mathbb{Z}_{2}$ monopoles in these theories
should be associated to weights in the coset with the highest weight
of the defining representation of $so(n)^{\vee}$. From these fundamental
$\mathbb{Z}_{2}$ monopoles we constructed other $\mathbb{Z}_{2}$
monopole asymptotic solutions. Differently from Weinberg's general
construction where the $\mathbb{Z}_{n}$ monopole's topological charge
sectors were associated to integers modulo $n$, in our construction
they are associated to cosets in the weight lattice $\Lambda_{\omega}(\widetilde{G}_{0}^{\vee})$,
which gives in principle a larger number of possible monopoles. We
construct the monopole solutions considering two symmetry breakings
of $su(n)$ to $so(n)$: one in which $so(n)$ is invariant under
outer automorphism and another in which it is invariant under Cartan
automorphism. In the first case, the monopole's magnetic flux is in
the Cartan subalgebra of $su(n)$ but $n$ must be odd, and in the
second case the magnetic flux is not in the Cartan subalgebra of $su(n)$.
This general procedure can be generalized to other gauge groups. We
expect that this explicit construction of the $\mathbb{Z}_{2}$ monopole
asymptotic solutions can be useful in order to understand better the
electromagnetic duality in the theories where the scalar field is
not in the adjoint representation.

This paper is organized as follows: we start in Sec. 2 giving some
mathematical conventions. Then, we give a brief review of the results
of GNO in Sec. 3 and explain in Sec. 4 how the topological charge
sectors of the $\mathbb{Z}_{n}$ monopoles are associated to particular
cosets in the weight lattice $\Lambda_{\omega}(\widetilde{G}_{0}^{\vee})$.
In Sec. 5, we obtain two scalar field configurations which break $su(n)$
to $so(n)$ where for the first configuration $so(n)$ is invariant
under Cartan automorphism and for the second configuration it is invariant
under outer automorphism. Finally, in Sec. 6 we construct the fundamental
$\mathbb{Z}_{2}$ monopoles for both symmetry breaks. We also include
an appendix where we analyze the center elements of $Spin(3),\, Spin(5),\, Spin(6)$.

\section{Mathematical conventions}

Let us start by giving some conventions which will be used later on.
Let $g$ be the Lie algebra of rank $r$ associated to the group%
\footnote{We shall adopt the convention of using capital letters to denote Lie
groups and lower letters for Lie algebras.%
} $G$. Let us adopt the Cartan-Weyl basis. In this basis, the commutation
relations read \begin{eqnarray}
\left[H_{i},E_{\alpha}\right] & = & \left(\alpha\right)^{i}E_{\alpha},\label{3.1}\\
\left[E_{\alpha},E_{-\alpha}\right] & = & \frac{2\alpha}{\alpha^{2}}\cdot H,\nonumber \end{eqnarray}
where generators $H_{i},\, i=1,\,2,\,...,\, r$, form a basis for
the Cartan subalgebra (CSA) $h$, $\alpha$ are roots and the upper
index in $\left(\alpha\right)^{i}$ means the component $i$ of $\alpha$. 

Given a representation of $g$, we can take a basis \{$\left|\mu\right\rangle $\}
in which the elements of the CSA, $H_{i},\, i=1,\,2,\,...,\, r$,
are diagonal,\[
H_{i}\left|\mu\right\rangle =\left(\mu\right)^{i}\left|\mu\right\rangle ,\,\,\, i=1,\,2,\,...,\, r.\]
The vector $\mu$ with the $r$ eigenvalues $\left(\mu\right)^{i}$
as components is called weight and $\left|\mu\right\rangle $ is called
weight state.

We shall denote by $\alpha_{i},\, i=1,2,...,r$, the simple roots
of $g$ which is a basis of the root space and by $\lambda_{i}$,
$i=1,2,...,r$, the fundamental weights of $g$. Moreover we shall
call \begin{equation}
\alpha_{i}^{\vee}=\frac{2\alpha_{i}}{\alpha_{i}^{2}},\,\,\,\,\,\lambda_{i}^{\vee}=\frac{2\lambda_{i}}{\alpha_{i}^{2}}\label{3.1a}\end{equation}
the simple coroots and fundamental coweights respectively. They satisfy
the relations\begin{equation}
\alpha_{i}\cdot\lambda_{j}^{\vee}=\alpha_{i}^{\vee}\cdot\lambda_{j}=\delta_{ij}.\label{3.1b}\end{equation}
$\alpha_{i}^{\vee}$ and $\lambda_{i}^{\vee}$ are respectively simple
roots and fundamental weights of the dual algebra $g^{\vee}$. 

Let us denote by $\widetilde{G}$ the covering group of $G$. Then,
the fundamental weights form a basis for the weight lattice of $\widetilde{G}$,\begin{equation}
\Lambda_{w}(\widetilde{G})=\left\{ \mu=\sum_{i=1}^{r}n_{i}\lambda_{i},\,\,\,\,\,\,\, n_{i}\in\mathbb{Z}\right\} .\label{3.3a}\end{equation}
This lattice includes as a subset, the root lattice of $G$, \begin{equation}
\Lambda_{r}(G)=\left\{ \beta=\sum_{i=1}^{r}n_{i}\alpha_{i},\,\,\,\,\,\,\, n_{i}\in\mathbb{Z}\right\} ,\label{3.3b}\end{equation}
which has the simple roots $\alpha_{i}$ as the basis. Similarly,
the fundamental coweights $\lambda_{i}^{\vee}$ are the basis of the
weight lattice of the dual group $\widetilde{G}^{\vee}$\begin{equation}
\Lambda_{w}(\widetilde{G}^{\vee})=\left\{ \mu=\sum_{i=1}^{r}n_{i}\lambda_{i}^{\vee},\,\,\,\,\,\,\, n_{i}\in\mathbb{Z}\right\} \label{3.3c}\end{equation}
which is also called the coweight lattice of $\widetilde{G}$ and
which has the root lattice of the dual group $G^{\vee}$(or coroot
lattice of $G$)\begin{equation}
\Lambda_{r}(G^{\vee})=\left\{ \beta=\sum_{i=1}^{r}n_{i}\alpha_{i}^{\vee},\,\,\,\,\,\,\, n_{i}\in\mathbb{Z}\right\} \label{3.3d}\end{equation}
as a subset.

\section{Magnetic monopoles in non-Abelian theories}

In a theory with gauge group $G$ spontaneously broken to $G_{0}$,
the monopole's solutions are associated to elements of the second
homotopy group\begin{equation}
\pi_{2}\left(G/G_{0}\right)=\textrm{Ker}\left(\pi_{1}\left(G_{0}\right)\rightarrow\pi_{1}\left(G\right)\right).\label{3.01}\end{equation}
This result implies that monopoles are associated with nontrivial
elements of $\pi_{1}\left(G_{0}\right)$ which correspond to trivial
elements of $\pi_{1}\left(G\right)$. Therefore, the relation (\ref{3.01})
is equivalent to \cite{BaisZ_2mono}\cite{Weinberg Book} \[
\pi_{2}\left(\widetilde{G}/G'_{0}\right)=\pi_{1}\left(G'_{0}\right)\]
where $G'_{0}$ is the unbroken subgroup of $\widetilde{G}$. Therefore
for simplicity, without loss of generality, we shall consider that
$G$ is simply connected.

Let us therefore consider a Yang-Mills-Higgs theory with gauge group
$G$ which we shall consider to be simple and simply connected. Let
us also consider that in the theory there is a scalar $\phi$ in a
representation $R$ of $G$ and $\phi_{0}$ is a vacuum configuration
which spontaneously breaks $G$ to a subgroup $G_{0}$, in such a
way that $\pi_{2}(G/G_{0})$ is nontrivial, which allows the existence
of magnetic monopoles. The generators of $G_{0}$ are those which
annihilate $\phi_{0}$, that is,\[
T_{a}\phi_{0}=0.\]
We shall denote by $g_{0}$ the algebra formed by these generators.

Let us briefly review some general properties of these monopoles which
will also be useful to fix our notation. Following GNO \cite{GNO},
we shall consider static finite energy monopoles with the asymptotic
form of the magnetic field of the form\begin{equation}
B_{i}(\theta,\varphi)=\frac{r_{i}}{4\pi r^{3}}X(\theta,\varphi)\,,\,\,\textrm{with}\,\,\,\, D_{i}X(\theta,\varphi)=0,\label{eq:3.1}\end{equation}
where $D_{i}\equiv\partial_{i}+ieW_{i}$ and $\theta,\varphi$ are
the angular spherical coordinates. The finite energy asymptotic condition
$D_{i}\phi=0$, implies that asymptotically we can write \cite{GNO}\begin{equation}
\phi(\theta,\varphi)=g(\theta,\varphi)\phi_{0},\label{5.0.2}\end{equation}
where $g(\theta,\varphi)\in G$. Then, the condition that $D_{i}X(\theta,\varphi)=0$
implies that \begin{equation}
X(\theta,\varphi)=g(\theta,\varphi)X_{0}g(\theta,\varphi)^{-1}\label{eq:3.1d}\end{equation}
with $X_{0}\equiv X(\theta=0,\varphi=0)$. The asymptotic condition
$D_{i}\phi=0$ and the definition of the field strength as the commutator
of covariant derivatives, implies that asymptotically $B_{i}\phi=0$.
Then, using (\ref{eq:3.1}), (\ref{5.0.2}), and (\ref{eq:3.1d})
results that $X_{0}\phi_{0}=0$ and therefore, $X_{0}\in g_{0}$.
Moreover, one can write \cite{GNO} \begin{equation}
X_{0}=\omega\cdot h\label{eq:3.3}\end{equation}
where $\omega\cdot h=\sum_{i}\omega_{i}h_{i}$, with $h_{i}$ being
the elements of the Cartan subalgebra (CSA) of $g_{0}$ and $\omega$
is a constant vector. Note that in general, the elements of the CSA
of $g_{0}$ do not necessarily belong to the CSA of $g$, the Lie
algebra of $G$. Therefore, we shall denote by $h_{i},\, f_{\alpha}$
the generators of $g_{0}$ and $H_{i}$ and $E_{\alpha}$ the generators
of $g$.

One can show the quantization condition \cite{GNO} \begin{equation}
\exp\left[ieX_{0}\right]=\exp\left[ie\omega\cdot h\right]=\mathbbm{1}.\label{3.32}\end{equation}
Let us consider that $G_{0}$ is semisimple%
\footnote{Note that if the scalar field $\phi$ is in the adjoint representation,
then $G_{0}$ is not semisimple since it has a $U(1)$ factor generated
by the vacuum configuration. This case is considered in detail in
\cite{Goddard and Olive papers}. %
}. In this case $G_{0}$ can be written as\[
G_{0}=\widetilde{G}_{0}/K(G_{0})\]
where $\widetilde{G}_{0}$ is the universal covering group of $G_{0}$
and the factor $K(G_{0})$ is the kernel of the homomorphism $\widetilde{G}_{0}\rightarrow G_{0}$.
The factor $K(G_{0})$ is a discrete subgroup of the center of $\widetilde{G}_{0}$
which we will denote by $Z(\widetilde{G}_{0}).$ Therefore, the topological
charge sectors of the theory are associated to \begin{equation}
\pi_{2}\left(G/G_{0}\right)=\pi_{1}\left(G_{0}\right)=K(G_{0})\subset Z(\widetilde{G}_{0}).\label{3.31}\end{equation}
Since $K(G_{0})$ is a cyclic group $\mathbb{Z}_{n}$ or a product
of cyclic groups, then these monopoles are called $\mathbb{Z}_{n}$
monopoles. 

Considering the condition (\ref{3.32}) in $\widetilde{G}_{0}$ rather
than in $G_{0}$ implies that \cite{GNO}\begin{equation}
\widetilde{\exp}\,\left[ie\omega\cdot h\right]\in K(G_{0})\subset Z(\widetilde{G}_{0}),\label{eq:3.5}\end{equation}
where $\widetilde{\exp}$ denotes the exponential mapping in $\widetilde{G}_{0}$.
Using the fact that the elements of the center $Z(\widetilde{G}_{0})$
of a group $G_{0}$, have the form\begin{equation}
\exp\left[2\pi iv\cdot h\right]\label{3.4}\end{equation}
where $v$ is a vector of the coweight lattice $\Lambda_{w}(\widetilde{G}_{0}^{\vee})$,
Goddard, Nuyts, and Olive concluded that the so-called magnetic weights
must satisfy \[
e\omega/2\pi\in\Lambda_{w}(\widetilde{G}_{0}^{\vee})\]
together with condition (\ref{eq:3.5}). From this result they conjectured
that the monopoles should be dual to particles in a theory with unbroken
gauge group $\widetilde{G}_{0}^{\vee}$.

\section{Topological charge sectors of $\mathbb{Z}_{n}$ monopoles}

Let us now analyze how the different values of $e\omega/2\pi$ are
associated to the different elements of $K(G_{0})\subset Z(\widetilde{G}_{0})$
or topological charge sectors (\ref{3.31}) of the theory. We shall
also restrict the possible values $e\omega/2\pi$ can take. In order
to do this we must remember that since $\Lambda_{r}(G_{0}^{\vee})$
is a sublattice of $\Lambda_{w}(\widetilde{G}_{0}^{\vee})$, we can
define the quotient $\Lambda_{w}(\widetilde{G}_{0}^{\vee})/\Lambda_{r}(G_{0}^{\vee})$.
This quotient can be represented by the cosets \cite{OliveTurok1}\begin{equation}
\Lambda_{r}(G_{0}^{\vee}),\,\,\,\,\,\lambda_{\tau(0)}^{\vee}+\Lambda_{r}(G_{0}^{\vee}),\,\,\,\,\,\lambda_{\tau^{2}(0)}^{\vee}+\Lambda_{r}(G_{0}^{\vee}),\,...\,,\,\,\,\,\lambda_{\tau^{n}(0)}^{\vee}+\Lambda_{r}(G_{0}^{\vee})\label{3.6}\end{equation}
where the weights $\lambda_{\tau^{q}(0)}^{\vee}$ are associated to
nodes of an extended Dynkin diagram of $G_{0}^{\vee}$ related to
the node $0$ by a symmetry transformation. In Table 1 we used black
nodes to denote these nodes in the extended Dynkin diagrams. One can
then show that the center of $G_{0}$ is a discrete group isomorphic
to the classes\cite{OTW2} \begin{eqnarray}
Z(\widetilde{G}_{0}) & = & \left\{ \exp\left[2\pi i\Lambda_{r}(G_{0}^{\vee})\cdot h\right],\,\exp\left[2\pi i\left(\lambda_{\tau(0)}^{\vee}+\Lambda_{r}(G_{0}^{\vee})\right)\cdot h\right],\,...\,,\right.\label{3.65}\\
 &  & \left.,\,...\,,\,\exp\left[2\pi i\left(\lambda_{\tau^{n}(0)}^{\vee}+\Lambda_{r}(G_{0}^{\vee})\right)\cdot h\right]\right\} .\nonumber \end{eqnarray}
 In other words, all the group elements (\ref{3.4}) with the vector
$v$ in the same coset of (\ref{3.6}) correspond to the same element
of the center $Z(\widetilde{G}_{0})$. In particular when $v$ belongs
to $\Lambda_{r}(G_{0}^{\vee})$, the group elements (\ref{3.4}) correspond
to the identity of $Z(\widetilde{G}_{0})$. Such a way of writing
the center elements is also quite useful to analyze the $\mathbb{Z}_{N}$
string solutions which appear when the gauge group $G$ is broken
to its center group $Z(G)$\cite{Z(N) strings}\cite{casimir law}\cite{string integrability}.
In \cite{casimir law} the center group of some groups is analyzed
in some more detail.

\begin{table}
\begin{center}\begin{tabular}{|c|c|c|c|}
\hline 
G&
Extended Dynkin diagram of $g$&
$W_{0}$&
$Z(G)$\tabularnewline
\hline
\hline 
$SU(n+1),\,\, n\geq2$&
\includegraphics[%
  bb=0bp 0bp 331bp 128bp,
  scale=0.3]{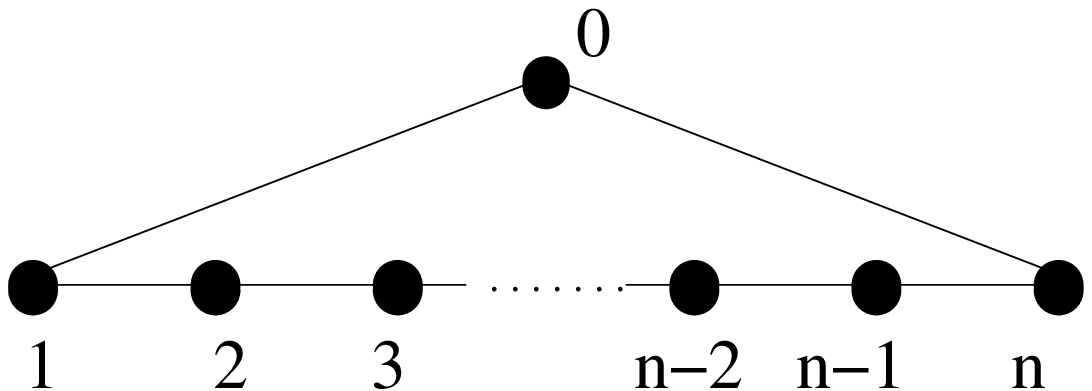}&
$0,1,2,...,n$&
$\mathbb{Z}_{n+1}$\tabularnewline
\hline 
$Spin(2n+1),\,\, n\geq3$&
\includegraphics[%
  bb=0bp 0bp 305bp 128bp,
  scale=0.3]{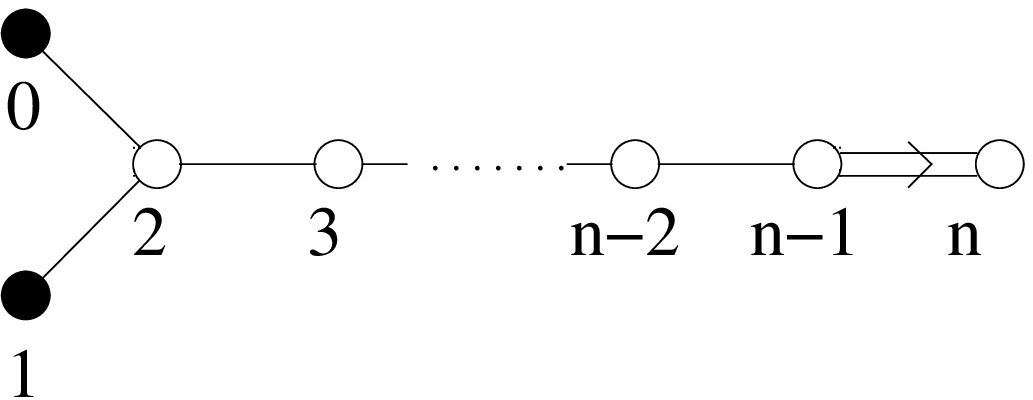}&
$0,1$&
$\mathbb{Z}_{2}$\tabularnewline
\hline 
$Sp(2n),\,\, n\geq2$&
\includegraphics[%
  bb=0bp 0bp 319bp 67bp,
  scale=0.3]{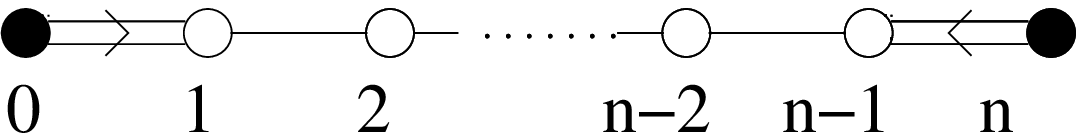}&
$0,n$&
$\mathbb{Z}_{2}$\tabularnewline
\hline 
$Spin(4n),\,\, n\geq2$&
\includegraphics[%
  bb=0bp 0bp 328bp 129bp,
  scale=0.3]{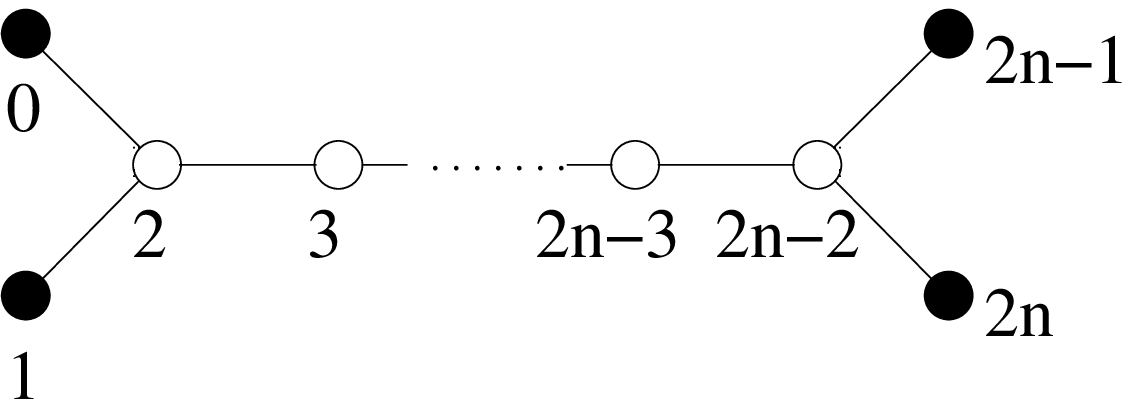}&
$0,1,2n-1,2n$&
$\mathbb{Z}_{2}\times\mathbb{Z}_{2}$\tabularnewline
\hline
$Spin(4n+2),\,\, n\geq2$&
\includegraphics[%
  bb=0bp 0bp 332bp 133bp,
  scale=0.3]{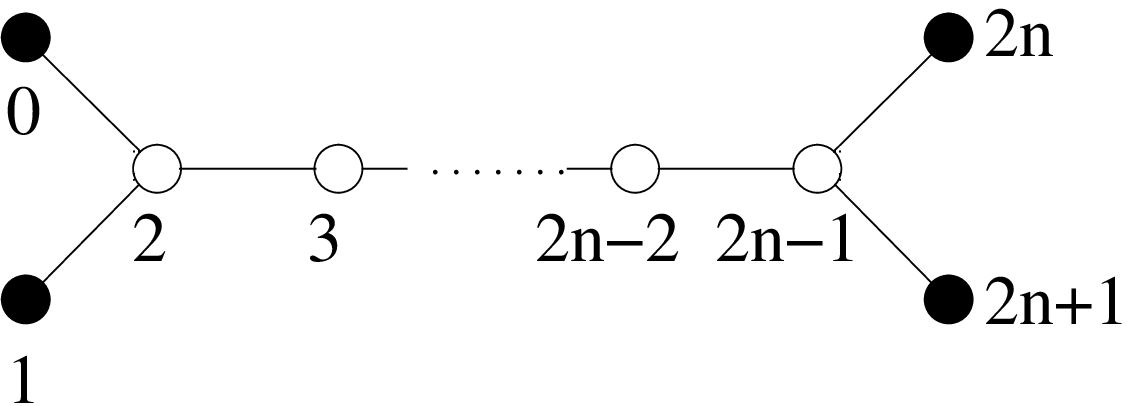}&
$0,1,2n,2n+1$&
$\mathbb{Z}_{4}$\tabularnewline
\hline
$E_{6}$&
\includegraphics[%
  bb=0bp 0bp 232bp 160bp,
  scale=0.3]{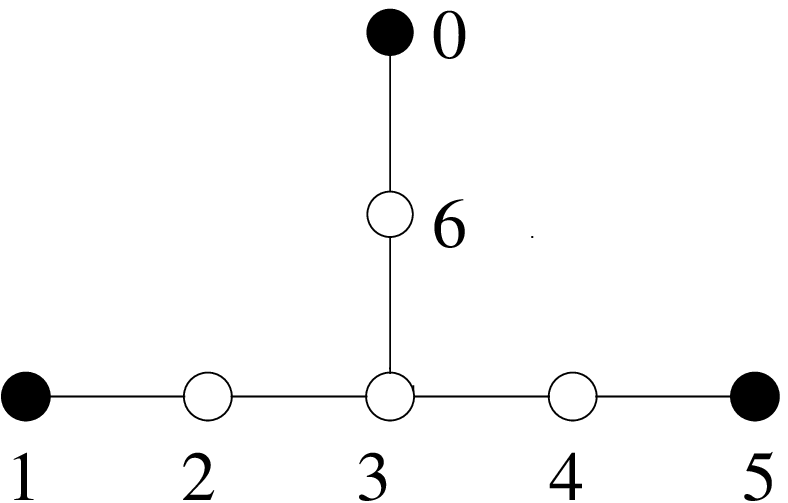}&
$0,1,5$&
$\mathbb{Z}_{3}$\tabularnewline
\hline 
$E_{7}$&
\includegraphics[%
  bb=0bp 0bp 394bp 105bp,
  scale=0.3]{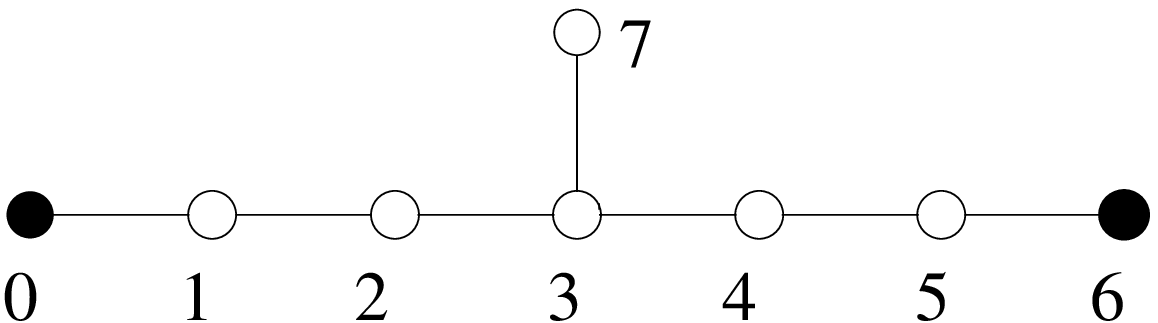}&
$0,6$&
$\mathbb{Z}_{2}$\tabularnewline
\hline
$E_{8}$&
\includegraphics[%
  bb=0bp 0bp 394bp 105bp,
  scale=0.3]{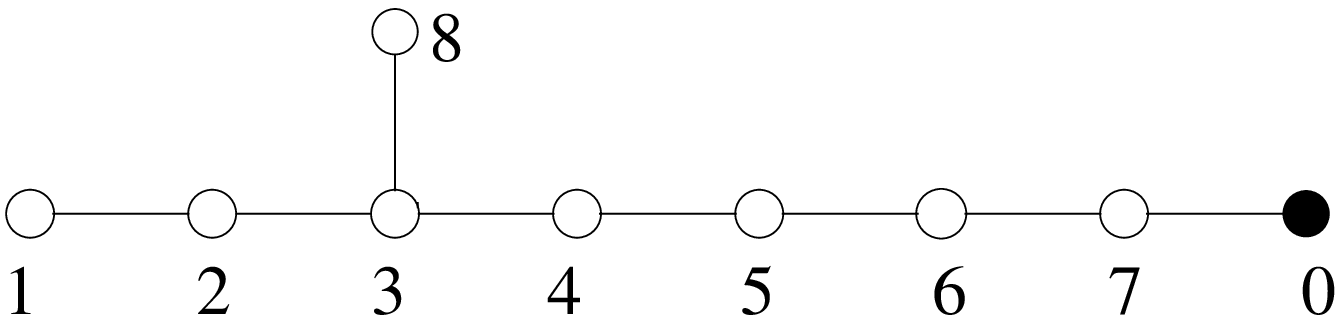}&
$0$&
$\mathbbm{1}$\tabularnewline
\hline 
$F_{4}$&
\includegraphics[%
  bb=0bp 0bp 232bp 66bp,
  scale=0.3]{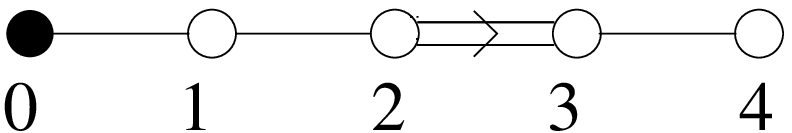}&
$0$&
$\mathbbm{1}$\tabularnewline
\hline 
$G_{2}$&
\includegraphics[%
  bb=0bp 0bp 122bp 70bp,
  scale=0.3]{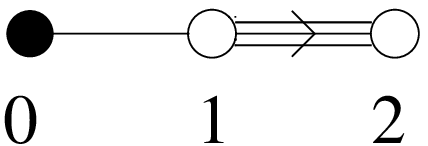}&
$0$&
$\mathbbm{1}$\tabularnewline
\hline
\end{tabular}\end{center}

\caption{\label{cap:Extended-Dynkin}\label{Dynkin}Extended Dynkin diagrams,
nodes symmetrically related to the node $0$ and center groups $Z(G)$. }
\end{table}

Since $K(G_{0})\subset Z(\widetilde{G}_{0})$, it will be formed by
a subset of elements of (\ref{3.65}). Therefore, the magnetic weights
$e\omega/2\pi$ cannot belong to an arbitrary coset of (\ref{3.6}),
but only to those cosets associated to elements of $K(G_{0})$, the
kernel of the homomorphism $\widetilde{G}_{0}\rightarrow G_{0}$.
Moreover, $\mathbb{Z}_{n}$ monopoles will be in the same topological
sector if their associated magnetic weights are in the same coset.
In the next sections we will analyze the values the magnetic weights
can take when the gauge group $SU(n)$ is broken to $Spin(n)/\mathbb{Z}_{2}.$

\section{Gauge symmetry breaking }

Let us now consider a Yang-Mills theory with gauge group $SU(n)$
and with a scalar field $\phi$ in the representation which is the
direct product $n\times n$ of $SU(n)$. In this case the theory can
in principle be embedded in a (deformed) ${\cal N}=2$ $SU(n)$ Super
Yang-Mills with an hypermultiplet in the $n\times n$ representation
which has a vanishing $\beta$ function. Such a theory has already
been considered in \cite{K x Z(k) theories} but with a different
gauge symmetry breaking, which gave rise to $\mathbb{Z}_{k}$ strings
and $\mathbb{Z}$ monopole confinement. We can also consider that
$\phi$ is in the symmetric part of $n\times n$, since the vacuum
solutions $\phi_{0}$ which we will consider are nontrivial only in
the symmetric part of $n\times n$. The specific form of the potential
is not important to determine the asymptotic form of the $\mathbb{Z}_{2}$
monopoles.

In order for $\mathbb{Z}_{2}$ monopoles to exist, we want to find
configurations of the scalar field which break $SU(n)$ to $Spin(n)/\mathbb{Z}_{2}$,
$n\geq3$. We first analyze some $so(n)$ subalgebras of $su(n)$.
We shall consider that $n\neq4$, since $so(4)=su(2)\oplus su(2)$
is not simple. We shall consider $so(n)$ invariant subalgebras under
Cartan or outer automorphisms which are order two automorphisms or
involutions. Remembering that, under an order two automorphism $\sigma$
of a Lie algebra $g$, $\sigma$ has eigenvalues $\exp(\pi ip),\, p=0,1$
and the Lie algebra split into $g^{(0)}$ and $g^{(1)}$, where $g^{(p)}$
is formed by the generators $T_{a}$, such that $\sigma(T_{a})=e^{\pi ip}T_{a},\,\, p=0,1$.
Moreover, $g^{(0)}$ forms a subalgebra of $g$. The quotient of the
group generated by $g$ modulo the group generated by $g^{(0)}$ is
a symmetric space and $g^{(1)}$ is associated to a representation
of $g^{(0)}$\cite{Helgason}.

\subsection{Breaking of su(n) to so(n) invariant under Cartan automorphism}

The Cartan automorphism for a general Lie algebra $g$ is defined
by\begin{eqnarray*}
\sigma\left(H_{i}\right) & = & -H_{i},\\
\sigma\left(E_{\alpha}\right) & = & -E_{-\alpha}.\end{eqnarray*}
For this automorphism, $g^{(0)}$ and $g^{(1)}$ are formed by the
generators \begin{eqnarray}
g^{(0)} & = & \left\{ E_{\alpha}-E_{-\alpha},\,\alpha>0\right\} ,\label{1.1}\\
g^{(1)} & = & \left\{ H_{a},\, a=1,2,...,\textrm{rank$(g)$};\, E_{\alpha}+E_{-\alpha},\,\alpha>0\right\} .\end{eqnarray}

Let us consider $g=su(n)$. {\LARGE }The generators of $su(n)$ in
the $n$-dimensional representation can can be written in terms of
the $n\times n$ matrices $E_{ij}$ with components $\left(E_{ij}\right)_{kl}=\delta_{ik}\delta_{jl}$
or \begin{equation}
E_{ij}\left|e_{j}\right\rangle =\left|e_{i}\right\rangle \label{1.5}\end{equation}
where $\left|e_{i}\right\rangle $ are the weight states of the $n$-dimensional
representation. Then, the basis elements of the CSA of $su(n)$ correspond
to the traceless combinations $E_{jj}-E_{j+1,j+1},$$j=1,2,\,...\,,n-1$.
On the other hand, the generator $E_{ij},\, i\neq j,$ is the step
operator associated to the root $e_{i}-e_{j}$, where $e_{i}$ are
orthonormal vectors in the $n$-dimensional vector space with $\left(e_{i}\right)_{k}=\delta_{ik}$.
The root $e_{i}-e_{j}$ is positive (negative) if $i<j$ $(i>j)$
and is a simple root if $j=i+1.$ The weight associated to $\left|e_{i}\right\rangle $
can be written as\[
e_{i}-\frac{1}{n}\sum_{j=1}^{n}e_{j}.\]
From (\ref{1.1}), we can conclude that the generators of $g^{(0)}$,
in the $n$-dimensional representation are\begin{equation}
M_{ij}=-i\left(E_{ij}-E_{ji}\right)\,,\,\,\, i<j,\label{1.6}\end{equation}
which are $n(n-1)$ antisymmetric $n\times n$ matrices which form
a $so(n)$ subalgebra of $su(n)$. Therefore, for $g=su(n),$ $g^{(0)}=so(n)$
for $n\ge3$\cite{Helgason}.

For example, in the $su(3)$ case, the Gell-Mann matrices \begin{eqnarray*}
\lambda_{2} & = & -i\left(E_{12}-E_{21}\right)=-i\left(E_{\alpha_{1}}-E_{-\alpha_{1}}\right),\\
\lambda_{5} & = & -i\left(E_{13}-E_{31}\right)=-i\left(E_{\alpha_{1}+\alpha_{2}}-E_{-\alpha_{1}-\alpha_{2}}\right),\\
\lambda_{7} & = & -i\left(E_{23}-E_{32}\right)=-i\left(E_{\alpha_{2}}-E_{-\alpha_{2}}\right)\end{eqnarray*}
form a $so(3)$ invariant subalgebra under Cartan automorphism. 

Let us consider the scalar field configuration\begin{equation}
\phi_{0}=v\sum_{i=1}^{n}\left|e_{i}\right\rangle \otimes\left|e_{i}\right\rangle \label{eq:1.2}\end{equation}
where $v$ is a constant. Using the fact that in a tensor product
representation a generator $T$ acts as $T\otimes\mathbbm{1}+\mathbbm{1}\otimes T$,
it is straightforward to conclude that $\phi_{0}$ is annihilated
only by the generators (\ref{1.6}) and hence \textit{breaks} $su(n)$
to $so(n)$ subalgebra generated by (\ref{1.6}). Therefore, we shall
consider that $\phi_{0}$ is the vacuum configuration responsible
for the symmetry breaking.

Let us determine an orthogonal basis of the Cartan subalgebra (CSA)
of the above $so(n)$ invariant subalgebra (\ref{1.6}). For $so(3)$,
we can consider the Gell-Mann matrix $\lambda_{2}$ as the generator
of the Cartan subalgebra. Recalling that $so(2m)$ and $so(2m+1)$
have same rank equal to $m$, one can check that the generators \begin{equation}
h_{k}=-i\left(E_{\alpha_{2k-1}}-E_{-\alpha_{2k-1}}\right),\,\,\,\, k=1,2,....,m,\label{eq:1.2a}\end{equation}
form an orthogonal basis of the Cartan subalgebras of $so(n)$ for
$n=2m,\,2m+1$, and where $E_{\alpha_{k}}$ are generators of $su(n)$.
It is important to note that for this $so(n)$ subalgebra invariant
under Cartan automorphism, the CSA of $so(n)$ is not in the CSA of
$su(n)$. We shall denote by $h_{i}$, $f_{\alpha}$ the generators
of the subalgebra $so(n)$ in order to distinguish from generators
$H_{i}$, $E_{\alpha}$ of $su(n)$.

\subsection{Breaking of su(2m+1) to so(2m+1) invariant under outer automorphism}

A Dynkin diagram which is invariant under a transformation of the
nodes, $i\rightarrow\tau(i)$, implies that the corresponding Cartan
matrix satisfies\[
K_{\tau(i)\tau(j)}=K_{ij}.\]
 As a consequence the associated Lie algebra $g$ has an outer automorphism%
\footnote{For a review see for example \cite{Helgason,OliveTurok1}.%
} $\tau$:\begin{eqnarray*}
\tau(\alpha\cdot H) & = & \tau\left(\alpha\right)\cdot H,\\
\tau\left(E_{\alpha}\right) & = & \chi_{\alpha}E_{\tau(\alpha)},\end{eqnarray*}
where $\chi_{\alpha}=\pm1$.

In particular, for $su(2m+1)$, the Dynkin diagram is invariant under
the transformation of the nodes $j\rightarrow2m+1-j$. Let $\left\langle j\right\rangle $
denote the orbit of nodes $j$ and $2m+1-j$ connected under this
transformation. Using the so-called folding procedure one can show
that the invariant subalgebra of $su(2m+1)$ under this automorphism
is $so(2m+1)$ \cite{OliveTurok1,Helgason}. {\LARGE }Let $H_{i}$
and $E_{\alpha_{i}}$ be generators of $su(2m+1)$. Then, the invariant
$so(2m+1)$ subalgebra has the following generators\begin{eqnarray}
H_{\left\langle l\right\rangle } & = & \sum_{i\in\left\langle l\right\rangle }\alpha_{i}\cdot H=H_{\alpha_{l}}+H_{\alpha_{2m+1-l}},\,\,\,\,\textrm{for }l=1,\,2,\,...,\, m-1,\nonumber \\
H_{\left\langle m\right\rangle } & = & 2\sum_{i\in\left\langle m\right\rangle }\alpha_{i}\cdot H=2\left(H_{\alpha_{m}}+H_{\alpha_{m+1}}\right),\label{eq:1.3}\\
E_{\pm\left\langle l\right\rangle } & = & \sum_{i\in\left\langle l\right\rangle }E_{\pm\alpha_{i}}=E_{\pm\alpha_{l}}+E_{\pm\alpha_{2m+1-l}},\,\,\,\,\textrm{for }l=1,\,2,\,...,\, m-1,\nonumber \\
E_{\pm\left\langle m\right\rangle } & = & \sqrt{2}\sum_{i\in\left\langle m\right\rangle }E_{\pm\alpha_{m}}=\sqrt{2}\left(E_{\pm\alpha_{m}}+E_{\pm\alpha_{m+1}}\right),\nonumber \end{eqnarray}
where $H_{\alpha}=2\alpha\cdot H/\alpha^{2}$.

Let $\alpha_{i}$ and $\lambda_{i}$ be, respectively, simple roots
and fundamental weights of $su(2m+1)$. Then, the simple roots, simple
coroots, fundamental weights and coweights of the invariant subalgebra
$so(2m+1)$ are\begin{eqnarray}
\alpha_{\left\langle l\right\rangle } & = & \frac{1}{2}\left(\alpha_{l}+\alpha_{2m+1-l}\right),\,\,\,\textrm{for }l=1,\,2,\,...,\, m\nonumber \\
\alpha_{\left\langle l\right\rangle }^{\vee} & = & \frac{2\alpha_{\left\langle l\right\rangle }}{\alpha_{\left\langle l\right\rangle }^{2}}=\alpha_{l}+\alpha_{2m+1-l},\,\,\,\textrm{for }l=1,\,2,\,...,\, m-1,\nonumber \\
\alpha_{\left\langle m\right\rangle }^{\vee} & = & \frac{2\alpha_{\left\langle m\right\rangle }}{\alpha_{\left\langle m\right\rangle }^{2}}=2\left(\alpha_{m}+\alpha_{m+1}\right),\label{4.2.3}\\
\lambda_{\left\langle l\right\rangle } & = & \frac{1}{2}\left(\lambda_{l}+\lambda_{2m+1-l}\right),\,\,\,\textrm{for }l=1,\,2,\,...,\, m-1,\nonumber \\
\lambda_{\left\langle m\right\rangle } & = & \frac{1}{4}\left(\lambda_{m}+\lambda_{m+1}\right),\nonumber \\
\lambda_{\left\langle l\right\rangle }^{\vee} & = & \frac{2\lambda_{\left\langle l\right\rangle }}{\alpha_{\left\langle l\right\rangle }^{2}}=\lambda_{l}+\lambda_{2m+1-l},\,\,\,\textrm{for }l=1,\,2,\,...,\, m.\nonumber \end{eqnarray}
One can check easily that the scalar products between the simple roots
give the Cartan matrix of $so(m+1)$ and that simple roots and fundamental
weights satisfy the right orthonormality conditions. 

As in the Cartan automorphism, we are looking for a vacuum configuration
$\phi_{0}$ which is annihilated by the generators given by (\ref{eq:1.3}),
that is, which breaks $su(2m+1)$ to the $so(2m+1)$ subalgebra invariant
by outer automorphism. Let us consider the scalar field configuration

\begin{equation}
\phi_{0}=v\sum_{l=1}^{2m+1}(-1)^{l+1}\left|e_{l}\right\rangle \otimes\left|e_{2m+2-l}\right\rangle ,\label{4.2.5}\end{equation}
 where $v$ is a constant. Since $\alpha_{l}+\alpha_{2m+1-l}=e_{l}-e_{l+1}+e_{2m+1-l}-e_{2m+2-l}$,
we can obtain directly that \[
H_{\left\langle l\right\rangle }\phi_{0}=0,\,\,\,\,\textrm{for}\,\, l=1,\,2,\,...\,,m.\]
 With respect to the folded step operators $E_{\left\langle l\right\rangle }$
we can use the fact that \[
E_{\alpha_{j}}=E_{j,j+1}\]
 which implies that \[
E_{\left\langle l\right\rangle }\phi_{0}=0,\,\,\,\,\textrm{for}\,\, l=1,\,2,\,...\,,m.\]
Therefore we can consider that $\phi_{0}$ given by Eq.(\ref{4.2.5})
is a configuration which breaks $su(2m+1)$ to the $so(2m+1)$ subalgebra
invariant by outer automorphism.

\subsection{Unbroken gauge group}

The above $so(n)$ subalgebras of $su(n)$ generates subgroups $G_{0}=Spin(n)/K(G_{0})$
of $SU(n)$, where $K(G_{0})$ is a subgroup of the center of $Spin(n)$
which we want to determine. 

Following \cite{Cornwell}, in order to determine the factor $K(G_{0})$
of the subgroup $G_{0}=\widetilde{G}_{0}/K(G_{0})$ of $G=\widetilde{G}/K(G)$,
we must first choose a representation $R_{\lambda}(\widetilde{G})$
of $\widetilde{G}$ with highest weight $\lambda$ such that the $\textrm{Ker}(R_{\lambda}(\widetilde{G}))=K(G)$.
If $R_{\lambda}(\widetilde{G})$ branches to the representation $R_{\overline{\lambda}}(\widetilde{G}_{0})$
of $\widetilde{G}_{0}$ with highest weight $\overline{\lambda}$,
then $K(G_{0})=\textrm{Ker(}R_{\overline{\lambda}}(\widetilde{G}_{0}))$. 

Therefore, in order to determine the discrete group $K(G_{0})$ of
the unbroken gauge subgroup $Spin(n)/K(G_{0})$ of $SU(n)$ we shall
choose the $n$-dimensional representation $R_{\lambda_{1}}(SU(n))$
of $SU(n)$ with highest weight $\lambda_{1}$ since in this representation
$\textrm{Ker}(R_{\lambda_{1}}(SU(n)))=\mathbbm{1}$ is well known.
Then, for the above two different embeddings of $so(n)$ in $su(n)$,
the $n$-dimensional irrep. of $su(n)$ branches to the $n$-dimensional
irreducible representation (irrep) of $so(n)$, which has $\lambda_{1}$
as highest weight. The weight states of this representation are of
the form $\left|\lambda_{1}-\gamma\right\rangle $, where $\gamma$
are positive roots of $so(n)$. The kernel $\textrm{Ker(}R_{\lambda_{1}}(Spin(n)))$
of this representation is made by the elements $g\in Spin(n)$ such
that\[
g\left|\lambda_{1}-\gamma\right\rangle =\left|\lambda_{1}-\gamma\right\rangle ,\]
for all weight states $\left|\lambda_{1}-\gamma\right\rangle $ of
the representation. Since $\textrm{Ker}(R_{\lambda_{1}}(Spin(n)))$
is a subgroup of the center $Z(Spin(n))$, we just need to act the
elements of $Z(Spin(n))$ on the weight states of the representation.
Let us consider $Spin(n)$ with $n\geq7$. In the Appendix we analyze
the particular cases of the groups $Spin(n)$ for $n=3,5,6$. From
the symmetry of the extended Dynkin diagram in Table 1 we can conclude
that the weight lattice $\Lambda_{w}(Spin(2n+1)^{\vee}),\, n\geq3$
split in\begin{equation}
\Lambda_{r}(Spin(2n+1)^{\vee}),\,\,\,\lambda_{1}^{\vee}+\Lambda_{r}(Spin(2n+1)^{\vee})\label{2.3}\end{equation}
and the center of $Spin(2n+1),\, n\geq3$ is\[
Z(Spin(2n+1))=\mathbb{Z}_{2}\cong\left\{ \exp\left(2\pi i\alpha^{\vee}\cdot h\right),\exp\left[2\pi i\left(\lambda_{1}^{\vee}+\alpha^{\vee}\right)\cdot h\right]\right\} ,\]
where $\alpha^{\vee}\in\Lambda_{r}(Spin(2n+1)^{\vee})$. Acting these
elements on the weight states of the $(2n+1)$-dimensional representation
of $so(2n+1)$ we obtain\begin{eqnarray}
\exp\left(2\pi i\alpha^{\vee}\cdot h\right)\left|\lambda_{1}-\gamma\right\rangle  & = & \left|\lambda_{1}-\gamma\right\rangle ,\label{2.311}\\
\exp\left[2\pi i\left(\lambda_{1}^{\vee}+\alpha^{\vee}\right)\cdot h\right]\left|\lambda_{1}-\gamma\right\rangle  & = & \left|\lambda_{1}-\gamma\right\rangle .\nonumber \end{eqnarray}

For $Spin(2n),$ we have\[
Z(Spin(2n))=\left\{ \begin{array}{ccl}
\mathbb{Z}_{2}\times\mathbb{Z}_{2} & \textrm{if} & 2n=4k,\\
\mathbb{Z}_{4} & \textrm{if} & 2n=4k+2,\end{array}\right.\]
where $k\in\mathbb{N}$. In both cases the weight lattice split in
the four cosets,\begin{equation}
\Lambda_{r}(Spin(2n)^{\vee}),\,\,\,\lambda_{1}^{\vee}+\Lambda_{r}(Spin(2n)^{\vee}),\,\,\,\lambda_{n-1}^{\vee}+\Lambda_{r}(Spin(2n)^{\vee}),\,\,\,\lambda_{n}^{\vee}+\Lambda_{r}(Spin(2n)^{\vee}),\label{5.12}\end{equation}
 and\begin{eqnarray*}
Z(Spin(2n)) & \cong & \left\{ \exp\left(2\pi i\alpha^{\vee}\cdot h\right),\exp\left[2\pi i\left(\lambda_{1}^{\vee}+\alpha^{\vee}\right)\cdot h\right],\right.\\
 &  & \hfill\left.\exp\left[2\pi i\left(\lambda_{n-1}^{\vee}+\alpha^{\vee}\right)\cdot h\right],\exp\left[2\pi i\left(\lambda_{n}^{\vee}+\alpha^{\vee}\right)\cdot h\right]\right\} ,\end{eqnarray*}
where $\alpha^{\vee}\in\Lambda_{r}(Spin(2n)^{\vee})$. Acting these
elements on the weight states of the $2n$-dimensional representation
of $so(2n)$ we obtain\begin{eqnarray}
\exp\left(2\pi i\alpha^{\vee}\cdot h\right)\left|\lambda_{1}-\gamma\right\rangle  & = & \left|\lambda_{1}-\gamma\right\rangle ,\nonumber \\
\exp\left[2\pi i\left(\lambda_{1}^{\vee}+\alpha^{\vee}\right)\cdot h\right]\left|\lambda_{1}-\gamma\right\rangle  & = & \left|\lambda_{1}-\gamma\right\rangle ,\label{2.312}\\
\exp\left[2\pi i\left(\lambda_{n-1}^{\vee}+\alpha^{\vee}\right)\cdot h\right]\left|\lambda_{1}-\gamma\right\rangle  & = & -\left|\lambda_{1}-\gamma\right\rangle ,\nonumber \\
\exp\left[2\pi i\left(\lambda_{n}^{\vee}+\alpha^{\vee}\right)\cdot h\right]\left|\lambda_{1}-\gamma\right\rangle  & = & -\left|\lambda_{1}-\gamma\right\rangle .\nonumber \end{eqnarray}
Therefore, from (\ref{2.311}) and (\ref{2.312}), we can conclude
that for $n$ odd or even, \begin{eqnarray*}
K(G_{0}) & = & \textrm{Ker}(R_{\lambda_{1}}(Spin(n))\\
 & = & \mathbb{Z}_{2}\cong\left\{ \exp\left[2\pi i\Lambda_{r}(Spin(n)^{\vee})\cdot h\right],\exp\left[2i\pi\left(\lambda_{1}^{\vee}+\Lambda_{r}(Spin(n)^{\vee})\right)\cdot h\right]\right\} \end{eqnarray*}
and the symmetry breaking is of the form\[
SU(n)\rightarrow Spin(n)/\mathbb{Z}_{2}\]

Hence we can conclude that for this symmetry breaking, the magnetic
weights $e\omega/2\pi$ must belong to the cosets\begin{equation}
\Lambda_{r}(Spin(n)^{\vee}),\,\,\,\lambda_{1}^{\vee}+\Lambda_{r}(Spin(n)^{\vee}),\label{4.3.5}\end{equation}
where $\lambda_{1}^{\vee}$ is the highest weight of the defining
representation of $so(n)^{\vee}$ which has dimension $2m$ for $so(2m)^{\vee}=so(2m)$
and for $so(2m+1)^{\vee}=sp(2m)$. This result holds also for the
special cases $Spin(3),\, Spin(5)$ and $Spin(6)$, as is analyzed
in the Appendix. A similar result was obtained in \cite{BaisZ_2mono}
using a different approach.

\section{$\mathbb{Z}_{2}$ monopole's asymptotic configuration}

We want to construct explicitly the asymptotic form for static spherically
symmetric $\mathbb{Z}_{2}$ monopole's solutions. In order to do that
let us define\begin{equation}
T_{3}^{\beta}=\frac{\beta\cdot h}{2},\label{eq:4.1}\end{equation}
where\[
\beta=\frac{e\omega}{2\pi}\in\Lambda_{\omega}(\widetilde{G}_{0}^{\vee})\]
and from Eq. (\ref{eq:3.5}) implies that\[
\widetilde{\exp}\,\left[4\pi iT_{3}^{\beta}\right]\in K(G_{0}).\]
Let us also consider other two generators $T_{1}^{\beta}$, $T_{2}^{\beta}$
of $g$, but not of $g_{0}$, such that \[
\left[T_{i}^{\beta},T_{j}^{\beta}\right]=i\epsilon_{ijk}T_{k}^{\beta}.\]
The choice of these two generators $T_{1}^{\beta}$ and $T_{2}^{\beta}$
will be discussed in detail in sections \ref{sub:so(n)-invariant-under}
and \ref{sub:so(2m+1)-invariant-under}. Following E. Weinberg et
al. \cite{WeinbergZ_2mono}, we can construct a spherically symmetric
monopole, consistent with GNO results, with the asymptotic form of
the scalar field given by (\ref{5.0.2}) with the group element $g(\theta,\varphi)$
of the form \[
g(\theta,\varphi)=\exp[-i\varphi T_{3}^{\beta}]\exp[-i\theta T_{2}^{\beta}]\exp[i\varphi T_{3}^{\beta}]\]
 and the asymptotic form for the gauge field given by \cite{WeinbergZ_2mono}\begin{equation}
W_{i}(\theta,\varphi)=g(\theta,\varphi)W_{i}^{0}g(\theta,\varphi)^{-1}+\frac{i}{e}\left(\partial_{i}g(\theta,\varphi)\right)g(\theta,\varphi)^{-1}\label{5.0.1}\end{equation}
where\begin{eqnarray*}
W_{r}^{0} & = & W_{\theta}^{0}=0,\\
W_{\phi}^{0} & = & \frac{T_{3}^{\beta}}{e}\left(1-\cos\theta\right).\end{eqnarray*}
This gauge field produces the magnetic field \begin{eqnarray}
B_{i}(\theta,\varphi) & = & \frac{r_{i}}{er^{3}}g(\theta,\varphi)T_{3}^{\beta}g(\theta,\varphi)^{-1}\nonumber \\
 & = & \frac{r_{i}}{4\pi r^{3}}g(\theta,\varphi)\,\omega\cdot h\, g(\theta,\varphi)^{-1}\label{5.0.4}\\
 & = & \frac{r_{i}}{4\pi r^{3}}X(\theta,\varphi),\nonumber \end{eqnarray}
consistent with Eqs. (\ref{eq:3.1}), (\ref{eq:3.1d}).

Note that in our construction, we have a difference from Weinberg's
construction. In his construction, the $\mathbb{Z}_{k}$ monopoles
in a given topological charge class were associated to the same integer
modulo $k$. On the other hand, in our construction, $\mathbb{Z}_{k}$
monopoles are in the same topological sector when they are associated
to magnetic weights $\beta=e\omega/2\pi$ in the same coset in (\ref{3.6}). 

Using the identity\begin{equation}
\exp\left(iaT_{j}\right)T_{i}\exp\left(-iaT_{j}\right)=\left(\cos\, a\right)T_{i}+\left(\sin\, a\right)\epsilon_{ijk}T_{k}\,,\,\,\,\, i\neq j\label{5.0.3}\end{equation}
where $a$ is a constant and $T_{i}$, $i=1,2,3$ are generators of
a $su(2)$ algebra, we can write the asymptotic form of the gauge
fields (\ref{5.0.1}) as\begin{eqnarray*}
W_{\theta}(\theta,\varphi) & = & -\frac{1}{e}\left[\left(\cos\varphi\right)T_{2}^{\beta}-\left(\sin\varphi\right)T_{1}^{\beta}\right],\\
W_{\varphi}(\theta,\varphi) & = & \frac{\sin\theta}{e}\left[-\left(\sin\theta\right)T_{3}^{\beta}+\cos\theta\left(\left(\cos\varphi\right)T_{1}^{\beta}+\left(\sin\varphi\right)T_{2}^{\beta}\right)\right],\\
W_{r}(\theta,\varphi) & = & 0.\end{eqnarray*}

Using Eq. (\ref{5.0.3}) we can also rewrite the $\mathbb{Z}_{2}$
monopole asymptotic magnetic field (\ref{5.0.4}) as \begin{eqnarray}
B_{i}(\theta,\varphi) & = & \frac{r_{i}}{er^{3}}\left[\left(\sin\theta\cos\varphi\right)T_{1}^{\beta}+\left(\sin\theta\sin\varphi\right)T_{2}^{\beta}+\left(\cos\theta\right)T_{3}^{\beta}\right]\nonumber \\
 & = & \frac{r_{i}}{er^{4}}\sum_{j=1}^{3}r_{j}T_{j}^{\beta},\label{5.0.6}\end{eqnarray}
which is the standard hedgehog form for the magnetic field.

Let us now analyze the possible monopole solutions for both symmetry
breakings discussed in the previous sections.

\subsection{so(n) invariant under Cartan automorphism\label{sub:so(n)-invariant-under}}

Let us determine the possible $su(2)$ subalgebra's generators $T_{i}^{\beta}$
for the symmetry breaking of $SU(n)\,\rightarrow\, Spin(n)/\mathbb{Z}_{2}$
where $Spin(n)$ is the subgroup invariant under Cartan automorphism.
We have that $T_{3}^{\beta}=\beta\cdot h/2$ where the Cartan elements
$h_{i}$ are given by (\ref{eq:1.2a}). Then, from Eq. (\ref{4.3.5}),
we conclude that for the $\mathbb{Z}_{2}$ monopoles associated to
the nontrivial sector, the vector $\beta$ must belong to the coset
\[
\lambda_{1}^{\vee}+\Lambda_{r}(Spin(n)^{\vee}).\]
This coset has in particular the weights of the defining representation
of the dual algebra $so(n)^{\vee}$ which has $\lambda_{1}^{\vee}$
as the highest weight. We know that $so(2m)^{\vee}=so(2m)$ and $so(2m+1)^{\vee}=sp(2m)$,
and that the weights of the defining representation of $so(2m)$ and
$sp(2m)$ have dimension $2m$. In terms of the orthonormal vectors
these weights can be written in both cases as \[
\pm e_{i},\, i=1,2,\,...,\, m.\]
For each weight $e_{k}$, we can construct a $su(2)$ subalgebra\begin{eqnarray}
T_{3}^{\pm e_{k}} & = & \pm\frac{1}{2}e_{k}\cdot h=\pm\frac{1}{2}h_{k}=\pm\frac{E_{\alpha_{2k-1}}-E_{-\alpha_{2k-1}}}{2i},\nonumber \\
T_{1}^{\pm e_{k}} & = & \frac{\alpha_{2k-1}\cdot H}{\alpha_{2k-1}^{2}},\label{6.1.1}\\
T_{2}^{\pm e_{k}} & = & \pm\frac{E_{\alpha_{2k-1}}+E_{-\alpha_{2k-1}}}{2},\nonumber \end{eqnarray}
for $k=1,2,...,m$. From Eq. (\ref{1.1}) we can conclude that $T_{3}^{\pm e_{k}}\in so(n)$
and $T_{1}^{\pm e_{k}},T_{2}^{\pm e_{k}}\notin so(n)$. Therefore,
for each weight of the defining representation of $so(n)^{\vee}$
we have a $\mathbb{Z}_{2}$ monopole solution (\ref{5.0.2}), (\ref{5.0.1}),(\ref{5.0.6}). 

We can construct monopole asymptotic forms with magnetic charge associated
to others elements of the cosets (\ref{4.3.5}). However, the $su(2)$
generators associated to these new monopoles seem to be always combination
of the generators (\ref{6.1.1}) and therefore these monopoles can
be interpreted as superpositions of the above monopoles which we call
fundamental. Some examples of these $su(2)$ subalgebras are \begin{eqnarray*}
T_{3}^{\pm n_{k}e_{k}} & = & \pm\sum_{k=1}^{m}\frac{1}{2}n_{k}e_{k}\cdot h=\pm\sum_{k=1}^{m}\frac{1}{2}n_{k}h_{k}=\pm\sum_{k=1}^{m}n_{k}\frac{E_{\alpha_{2k-1}}-E_{-\alpha_{2k-1}}}{2i},\\
T_{1}^{\pm n_{k}e_{k}} & = & \sum_{k=1}^{m}n_{k}\frac{\alpha_{2k-1}\cdot H}{\alpha_{2k-1}^{2}},\\
T_{2}^{\pm n_{k}e_{k}} & = & \pm\sum_{k=1}^{m}n_{k}\frac{E_{\alpha_{2k-1}}+E_{-\alpha_{2k-1}}}{2},\end{eqnarray*}
for $n_{k}=0,1$.

We can understand easily the $\mathbb{Z}_{2}$ nature of these monopoles:
writing the fundamental weight $\lambda_{1}$ in the basis of simple
roots, we have that for $sp(2m)=so(2m+1)^{\vee}$,\[
\lambda_{1}=\alpha_{1}+\alpha_{2}+\ldots+\alpha_{m-1}+\frac{1}{2}\alpha_{m}\]
and for $so(2m)=so(2m)^{\vee}$,\[
\lambda_{1}=\alpha_{1}+\alpha_{2}+\ldots+\alpha_{m-2}+\frac{1}{2}\left(\alpha_{m-1}+\alpha_{m}\right).\]
In both cases we see that $2\lambda_{1}\in\Lambda_{r}\left(Spin(n)^{\vee}\right)$,
and therefore a combination of an even number of fundamental $\mathbb{Z}_{2}$
monopoles will result in a configuration associated to $\Lambda_{r}\left(Spin(n)^{\vee}\right)$
which corresponds to the trivial element $\mathbbm{1}$ of the group
$\mathbb{Z}_{2}$, and an odd combination will result on a configuration
associated to the nontrivial coset, which corresponds to element $-\mathbbm{1}$
of the group $\mathbb{Z}_{2}$. We must also note that two configurations
belonging to the same topological sector does not have necessarily
the same magnetic charge or magnetic weight.

From the vacuum (\ref{eq:1.2}), we obtain that the asymptotic form
of the scalar field for the monopole associated to the magnetic weight
$\beta=\pm e_{k}$ is\begin{eqnarray*}
\phi(\theta,\varphi) & = & g(\theta,\varphi)\phi_{0}\\
 & = & \phi_{0}+v\left\{ \left(\cos\theta-1\right)\mp i\sin\theta\cos\varphi\right\} \left\{ \left|2k-1,2k-1\right\rangle +\left|2k,2k\right\rangle \right\} \\
 &  & \mp iv\sin\theta\cos\varphi\left\{ \left|2k,2k-1\right\rangle +\left|2k-1,2k\right\rangle \right\} ,\end{eqnarray*}
where we defined $\left|i,j\right\rangle =\left|e_{i}\right\rangle \otimes\left|e_{j}\right\rangle .$

From the kinetic term for the scalar field, expanding $\phi$ around
the vacuum $\phi_{0}$, we obtain the term 

\[
D_{\mu}\phi_{0}^{\dagger}D^{\mu}\phi_{0}=\frac{e^{2}}{2}\phi_{0}^{\dagger}\left\{ T_{a},T_{b}\right\} \phi_{0}W_{a\mu}W_{b}^{\mu},\]
 which implies that the mass squared matrix for the gauge particles
is\[
\left(M^{2}\right)_{ab}=e^{2}\phi_{0}^{\dagger}\left\{ T_{a},T_{b}\right\} \phi_{0}.\]
For $su(n)$ broken to $so(n)$, there are $(n+2)(n-1)/2$ massive
gauge particles which can be associated to the generators\begin{eqnarray*}
T_{1}^{ij} & = & \frac{E_{ij}+E_{ji}}{2},\,\,\, i<j,\,\,\,\, i=1,2,...\,,n-1,\\
T_{3}^{i,i+1} & = & \frac{E_{ii}-E_{i+1,i+1}}{2},\,\,\,\,\, i=1,2,...\,,n-1.\end{eqnarray*}
Then, using the definition of $E_{ij}$ and adopting the normalization
$\left\langle i,j\right|\left.k,l\right\rangle =\delta_{ik}\delta_{jl}$
one can obtain directly that all massive gauge particles have same
mass equal to \[
m=2ev.\]
This result coincides with the one obtained in \cite{London} for
the $SU(3)$ case, up to a global factor due to a different normalization.

\subsection{so(2m+1) invariant under outer automorphism\label{sub:so(2m+1)-invariant-under}}

Let us determine the possible $su(2)$ subalgebra generators $T_{i}^{\beta}$
we can have for the symmetry breaking of $SU(2m+1)\,\rightarrow\, Spin(2m+1)/\mathbb{Z}_{2}$
where $Spin(2m+1)$ is the subgroup invariant under outer automorphism.
From Eq. (\ref{4.3.5}), we can conclude that for the $\mathbb{Z}_{2}$
monopoles in the nontrivial sector, the vector $\beta$ must belong
to coset \[
\beta=\lambda_{\left\langle 1\right\rangle }^{\vee}+\sum_{i=1}^{m}c_{\left\langle i\right\rangle }\alpha_{\left\langle i\right\rangle }^{\vee},\]
where $c_{\left\langle i\right\rangle }$ are integer numbers and
$\lambda_{\left\langle i\right\rangle }^{\vee}$ and $\alpha_{\left\langle i\right\rangle }^{\vee}$
are, respectively, coweights and coroots of $so(2m+1)$ given by the
set of Eqs. (\ref{4.2.3}). From these equations and the fact that
\[
\lambda_{\left\langle 1\right\rangle }^{\vee}=\lambda_{1}+\lambda_{2m}=\psi=\alpha_{1}+\alpha_{2}+\ldots+\alpha_{2m}\]
where $\psi$ is the highest root of $su(2m+1)$, we can conclude
that\begin{equation}
\beta=\sum_{i=1}^{m}\left(1+c_{\left\langle i\right\rangle }\right)\left(\alpha_{i}+\alpha_{2m+1-i}\right).\label{5.1.4}\end{equation}
Therefore, $\beta$ must also belong to the subspace of $\Lambda_{r}(SU(2m+1))$
invariant under the outer automorphism transformation $\tau(\alpha_{i})=\alpha_{2m+1-i}$
of the $su(2m+1)$ algebra. The fact that at the same time $\beta\in\Lambda_{w}(Spin(2m+1)^{\vee})$
and $\beta\in\Lambda_{r}(SU(2m+1))$ is consistent with Eq. (\ref{3.01})
which means that the $\mathbb{Z}_{2}$ monopoles must be associated
to elements of $\pi_{1}\left(Spin(2m+1)/\mathbb{Z}_{2}\right)=\mathbb{Z}_{2}$
which correspond to the identity of $\pi_{1}\left(SU(2m+1)\right)=\mathbbm{1}$.
In order to construct $su(2)$ subalgebras we consider that $\beta$
must satisfy not only condition (\ref{5.1.4}) but also that it is
a root of $su(2m+1).$ Then we can define \begin{eqnarray}
T_{3}^{\beta} & = & \frac{\beta\cdot H}{2}=\frac{\beta\cdot H}{\beta^{2}},\nonumber \\
T_{1}^{\beta} & = & \frac{E_{\beta}+E_{-\beta}}{2},\label{5.1.5}\\
T_{2}^{\beta} & = & \frac{E_{\beta}-E_{-\beta}}{2i},\nonumber \end{eqnarray}
where we used the fact that $\beta^{2}=2$ since it is a root of $su(2m+1)$.
Since the roots of $su(2m+1)$ are of the form\[
\alpha_{p}+\alpha_{p+1}+\alpha_{p+2}+\ldots+\alpha_{p+q}\]
where $0\leq q\leq2m-p$, and $\beta$ must satisfy (\ref{5.1.4}),
we arrive to the conclusion that $\beta$ can be the following $2m$
roots,\begin{eqnarray}
\alpha_{1}+\alpha_{2}+\ldots+\alpha_{2m} & = & \lambda_{\left\langle 1\right\rangle }^{\vee},\nonumber \\
\alpha_{2}+\alpha_{3}+\ldots+\alpha_{2m-1} & = & \lambda_{\left\langle 1\right\rangle }^{\vee}-\alpha_{\left\langle 1\right\rangle }^{\vee},\nonumber \\
\vdots &  & \vdots\nonumber \\
\alpha_{m}+\alpha_{m+1} & = & \lambda_{\left\langle 1\right\rangle }^{\vee}-\alpha_{\left\langle 1\right\rangle }^{\vee}-\ldots-\alpha_{\left\langle m-1\right\rangle }^{\vee},\label{5.1.7}\\
-\left(\alpha_{m}+\alpha_{m+1}\right) & = & \lambda_{\left\langle 1\right\rangle }^{\vee}-\alpha_{\left\langle 1\right\rangle }^{\vee}-\ldots-\alpha_{\left\langle m-1\right\rangle }^{\vee}-\alpha_{\left\langle m\right\rangle }^{\vee},\nonumber \\
-\left(\alpha_{m-1}+\alpha_{m}+\alpha_{m+1}+\alpha_{m+2}\right) & = & \lambda_{\left\langle 1\right\rangle }^{\vee}-\alpha_{\left\langle 1\right\rangle }^{\vee}-\ldots-2\alpha_{\left\langle m-1\right\rangle }^{\vee}-\alpha_{\left\langle m\right\rangle }^{\vee},\nonumber \\
\vdots &  & \vdots\nonumber \\
-\left(\alpha_{1}+\alpha_{2}+\ldots+\alpha_{2m}\right) & = & \lambda_{\left\langle 1\right\rangle }^{\vee}-2\alpha_{\left\langle 1\right\rangle }^{\vee}-\ldots-2\alpha_{\left\langle m-1\right\rangle }^{\vee}-\alpha_{\left\langle m\right\rangle }^{\vee},\nonumber \end{eqnarray}
where we wrote the roots of $su(2m+1)$ as coweights of $so(2m+1)$,
using Eq. (\ref{4.2.3}). From the fact that fundamental coweights
and simple coroots of $so(2m+1)$ are, respectively, fundamental weights
and simple roots of $sp(2m),$ we can recognize this set as the weights
of the $2m$-dimensional defining representation of $sp(2m)$.

It remains to show that $T_{1}^{\beta},T_{2}^{\beta}\notin so(2m+1)$.
In order to do that we can write the set of roots (\ref{5.1.7}) in
terms of the orthonormal basis vectors $e_{i},\, i=1,2,\,\ldots,2m+1$.
Since $\alpha_{i}=e_{i}-e_{i+1}$, the roots in (\ref{5.1.7}) are
of the form \begin{equation}
e_{p}-e_{2m+2-p},\,\,\, p=1,2,\ldots,m,\label{5.1.8}\end{equation}
 consistent with the fact that under the outer automorphism of $su(2m+1),$$\tau(e_{p})=-e_{2m+2-p}$.
The step operator associated to the root (\ref{5.1.8}) in the $n$-dimensional
representation is proportional to the matrix $E_{p,2m+2-p}.$Therefore,
denoting \[
\mathbb{E}_{ij}=E_{ij}\otimes\mathbbm{1}+\mathbbm{1}\otimes E_{ij},\]
we can write $T_{1}^{\beta}$ and $T_{2}^{\beta}$ in terms of these
matrices and acting on the vacuum (\ref{4.2.5}) we obtain\begin{eqnarray*}
2T_{1}^{p,2m+2-p}\phi_{0} & = & \left(\mathbb{E}_{p,2m+2-p}-\mathbb{E}_{2m+2-p,p}\right)v\sum_{l=1}^{2m+1}(-1)^{l+1}\left|e_{l}\right\rangle \otimes\left|e_{2m+2-l}\right\rangle \\
 & = & 2(-1)^{p+1}v\left(\left|e_{p}\right\rangle \otimes\left|e_{p}\right\rangle +\left|e_{2m+2-p}\right\rangle \otimes\left|e_{2m+2-p}\right\rangle \right)\\
 & \neq & 0,\end{eqnarray*}
\begin{eqnarray*}
2iT_{2}^{p,2m+2-p}\phi_{0} & = & \left(\mathbb{E}_{p,2m+2-p}+\mathbb{E}_{2m+2-p,p}\right)v\sum_{l=1}^{2m+1}(-1)^{l+1}\left|e_{l}\right\rangle \otimes\left|e_{2m+2-l}\right\rangle \\
 & = & 2(-1)^{p+1}v\left(\left|e_{p}\right\rangle \otimes\left|e_{p}\right\rangle -\left|e_{2m+2-p}\right\rangle \otimes\left|e_{2m+2-p}\right\rangle \right)\\
 & \neq & 0.\end{eqnarray*}
Hence, $T_{1}^{\beta},T_{2}^{\beta}\notin so(2m+1)$. Therefore, we
can conclude that to each weight (\ref{5.1.7}) of the defining representation
of $sp(2m)=so(2m+1)^{\vee}$ we can associate a $su(2)$ subalgebra
(\ref{5.1.5}) and $\mathbb{Z}_{2}$ monopole. 

Similarly to the previous case, we can construct monopole asymptotic
forms with magnetic charge associated to others elements of the cosets
(\ref{4.3.5}). However, the $su(2)$ generators associated to these
new monopoles seem to always be combination of the generators (\ref{5.1.5})
and therefore these monopoles can be interpreted as superpositions
of the above monopoles. Some examples of these $su(2)$ subalgebras
are generated by\begin{eqnarray*}
T_{3}^{n_{p}(e_{p}-e_{2m+2-p})} & = & \sum_{p=1}^{m}n_{p}\frac{\left(e_{p}-e_{2m+2-p}\right)\cdot H}{2},\\
T_{1}^{n_{p}(e_{p}-e_{2m+2-p})} & = & \sum_{p=1}^{m}n_{p}\frac{E_{(e_{p}-e_{2m+2-p})}+E_{(e_{p}-e_{2m+2-p})}}{2},\\
T_{2}^{n_{p}(e_{p}-e_{2m+2-p})} & = & \sum_{p=1}^{m}n_{p}\frac{E_{(e_{p}-e_{2m+2-p})}-E_{(e_{p}-e_{2m+2-p})}}{2i},\end{eqnarray*}
where $n_{p}=0,1$.

For the vacuum (\ref{4.2.5}), we obtain that the asymptotic form
of the scalar field for the monopole associated to $\beta=e_{p}-e_{2m+2-p}$
is\begin{eqnarray*}
\phi(\theta,\varphi) & = & g(\theta,\varphi)\phi_{0}\\
 & = & \phi_{0}+(-1)^{p+1}v\left\{ -\sin\theta\left[e^{-i\varphi}\left|p,p\right\rangle +e^{i\varphi}\left|2m+2-p,2m+2-p\right\rangle \right]+\right.\\
 &  & +\left.\left(\cos\theta-1\right)\left[\left|p,2m+2-p\right\rangle +\left|2m+2-p,p\right\rangle \right]\right\} .\end{eqnarray*}

\section*{Appendix}

Let us analyze the elements of the center group and the kernel $K(G_{0})$
for the special cases of $Spin(n)$ for $n=3,5,6$: 

For $Spin(3)\cong SU(2)$, the weight lattice splits in two cosets,\[
\Lambda_{r}(SU(2)),\,\,\,\lambda_{1}^{\vee}+\Lambda_{r}(SU(2)),\]
and\[
Z(SU(2))=\mathbb{Z}_{2}\cong\left\{ \exp\left(2\pi i\alpha^{\vee}\cdot h\right),\exp\left[2\pi i\left(\lambda_{1}^{\vee}+\alpha^{\vee}\right)\cdot h\right]\right\} .\]
The branching of the irrep. with highest weight $\lambda_{1}$ of
$su(3)$ in $su(2)$ is $(1\,0)\simeq(2)$. Acting the center elements
on the weight states $\left|2\lambda_{1}-\gamma\right\rangle $ of
$su(2)$, we obtain that\[
\textrm{Ker}(R_{\lambda_{1}}(Spin(2)))=\mathbb{Z}_{2}\cong\left\{ \exp\left[\left(2\pi i\Lambda_{r}(SU(2))\right)\cdot h\right],\exp\left[2\pi i\left(\lambda_{1}^{\vee}+\Lambda_{r}(SU(2))\right)\cdot h\right]\right\} ,\]
where $\lambda_{1}^{\vee}$ is the highest weight of the $2$-dimensional
irrep. of $su(2)$.

For $Spin(5)\cong Sp(4)$, the weight lattice splits in two cosets,\[
\Lambda_{r}(Sp(4)^{\vee}),\,\,\,\lambda_{2}^{\vee}+\Lambda_{r}(Sp(4)^{\vee})\]
and\[
Z(Sp(4))=\mathbb{Z}_{2}\cong\left\{ \exp\left(2\pi i\alpha^{\vee}\cdot h\right),\exp\left[2\pi i\left(\lambda_{2}^{\vee}+\alpha^{\vee}\right)\cdot h\right]\right\} .\]
The branching of the irrep. with highest weight $\lambda_{1}$ of
$su(5)$ in $sp(4)$ is $(1\,0\,0\,0)\simeq(0\,1)$. Acting the center
elements on the weight states $\left|\lambda_{2}-\gamma\right\rangle $
of $sp(4)$, we obtain that\[
\textrm{Ker}(R_{\lambda_{2}}(Spin(5))=\mathbb{Z}_{2}\cong\left\{ \exp\left[\left(2\pi i\Lambda_{r}(Sp(4)^{\vee})\right)\cdot h\right],\exp\left[2\pi i\left(\lambda_{2}^{\vee}+\Lambda_{r}(Sp(4)^{\vee})\right)\cdot h\right]\right\} \]
where $\lambda_{2}^{\vee}$ is the highest weight of the $4$-dimensional
irrep. of $sp(4)$.

For $Spin(6)\cong SU(4)$, the weight lattice splits in four cosets,
\[
\Lambda_{r}(SU(4)),\,\,\,\lambda_{1}+\Lambda_{r}(SU(4)),\,\,\,\lambda_{2}+\Lambda_{r}(SU(4)),\,\,\,\lambda_{3}+\Lambda_{r}(SU(4)),\]
and\begin{eqnarray*}
Z(SU(4)) & \cong & \left\{ \exp\left(2\pi i\alpha^{\vee}\cdot h\right),\exp\left[2\pi i\left(\lambda_{1}^{\vee}+\alpha^{\vee}\right)\cdot h\right],\right.\\
 &  & \hfill\left.\exp\left[2\pi i\left(\lambda_{2}^{\vee}+\alpha^{\vee}\right)\cdot h\right],\exp\left[2\pi i\left(\lambda_{3}^{\vee}+\alpha^{\vee}\right)\cdot h\right]\right\} .\end{eqnarray*}
The branching of the irrep. with highest weight $\lambda_{1}$ of
$su(6)$ in $su(4)$ is $(1\,0\,0\,0\,0)\simeq(0\,1\,0)$. Acting
the center elements on the weight states $\left|\lambda_{2}-\gamma\right\rangle $
of $su(4)$, we obtain that\[
\textrm{Ker}(R_{\lambda_{1}}(Spin(6))=\mathbb{Z}_{2}\cong\left\{ \exp\left[\left(2\pi i\Lambda_{r}(SU(4))\right)\cdot h\right],\exp\left[2\pi i\left(\lambda_{2}^{\vee}+\Lambda_{r}(SU(4))\right)\cdot h\right]\right\} \]
where $\lambda_{2}^{\vee}$ is the highest weight of the $6$-dimensional
irrep. of $su(4)\cong so(6)$.

Therefore for all the three cases, similarly to the general case of
$Spin(n),\, n\geq7$, \[
\textrm{Ker}(R_{\lambda_{i}}(Spin(n))=\mathbb{Z}_{2}\cong\left\{ \exp\left[\left(2\pi i\Lambda_{r}(Spin(n)^{\vee})\right)\cdot h\right],\exp\left[2\pi i\left(\lambda_{i}^{\vee}+\Lambda_{r}(Spin(n)^{\vee})\right)\cdot h\right]\right\} ,\]
where $\lambda_{i}^{\vee}$ is the highest weight of the defining
representation of $so(n)^{\vee}$ which has dimension $2m$ for $so(2m)^{\vee}=so(2m)$
and for $so(2m+1)^{\vee}=sp(2m)$.

\subsection*{Acknowledgement}

M.A.C.K. wish to thank L.A. Ferreira and T. Hollowood for useful discussions
and ICTP, Italy, where this work was partly carried out. P.J.L. is
grateful to CNPq for financial support.

\end{document}